\documentstyle[12pt]{article}
\begin{document}
\title{GDH2000 Convenor's Report: Spin Polarizabilities}
\author{Barry R. Holstein\\
Department of Physics-LGRT\\
University of Massachusetts\\
Amherst, MA  01003}
\maketitle
\begin{abstract}
The subject of low energy polarized Compton scattering form the 
proton, which is characterized by phenomenological spin-polarizabilities, 
is introduced and connection is made to new theoretical and experimental 
developments which were reported to this meeting.
\end{abstract}
\newpage

\section{Introduction}
It is a pleasure to report on what was a very stimulating session involving 
low energy aspects of the GDH sum rule, which can be characterized in terms of 
so-called "spin-polarizabilities."  Earlier at this meeting 
we heard a quote from J\"{u}rgen
Ahrens, the essence of which was that details of the GDH integrand were of
more interest than the final number itself.  I concur with his statement and
in fact would add a corollary which asserts that the decomposition into 
intrinsic amplitudes is even more important than study of the cross section.
I think that this will become clear in the discussion below.

One aspect of this low energy analysis was stressed by G. Krein\cite{kre}, 
who pointed
out that the component of the GDH integrand studied experimentally at MAMI 
and reported at this workshop involves the energy range 200 MeV$\leq\omega\leq$
800 MeV and omits the region $\sim m_\pi\leq\omega\leq 200$ MeV.  
Thus in order to perform the
GDH integration between the inelastic threshold and 200 MeV requires 
some sort of reliable theoretical
input such as provided by the MAID or SAID analysis of low energy pion
photoproduction.  
In this regard, Krein emphasized that until
recently the charge pion $E_{0+}$ multipole, which dominates the low energy
cross section, was underpredicted by SAID by about 15\%, compared both 
to experiment
and to the Kroll-Ruderman stricture\cite{kr}.  To show the
importance of having the correct behavior in this near threshold region,
he noted that use of the SAID $E_{0+}$ multipole leads to a 20$\mu$b
($\sim$ 10\%) change in the GDH sum rule value and a $0.75\times
10^{-4}$ fm$^4$ ($\sim$ 100\%) shift in the value of the forward spin
polarizability, to be discussed below.

\section{Real Compton Scattering}

It is thus very important to have a proper multipole analysis of the low
energy Compton amplitude in order to perform a correct GDH analysis.  In the
process one can also learn a good deal about nucleon structure.  In order to 
see how this comes about, consider first {\it very} low energy Compton
scattering---say $\omega << 20$ MeV---wherein the photon wavelength is much
longer than the size of the nucleon.  In this case, one is unable
to resolve the structure of the target and is sensitive only to its overall
charge--$e$--and mass--$m.$  The interaction is described by the 
simple lowest order Hamiltonian
\begin{equation}
H={(\vec{p}-e\vec{A})^2\over 2m}+e\phi
\end{equation}
and the resultant Compton scattering amplitude has the canonical Thomson form
\begin{equation}
{\rm Amp}=-{e^2\over m}\hat{\epsilon}'\cdot\hat{\epsilon}
\end{equation}
At higher energy---shorter wavelength---the internal structure becomes visible and
one can describe the interaction in terms of an effective Hamiltonian
having certain elementary properties---
\begin{itemize}
\item [i)] quadratic in $\vec{A}$;
\item [ii)] gauge invariant;
\item [iii)] rotational scalar;
\item [iv)] P,T even, etc.
\end{itemize} 
To the next order then the resultant form of the interaction is unique
and must have the from
\begin{equation}
H_{eff}=-{1\over 2}4\pi\alpha_E\vec{E}^2-{1\over 2}4\pi\beta_M\vec{H}^2
\end{equation}
where $\alpha_E,\,\beta_M$ are phenomenological constants having the dimensions
of volume.  The physical meaning of these constants can be 
seen from the definitions
\begin{equation}
\vec{p}=-{\delta H_{eff}\over \delta\vec{E}}=4\pi\alpha_E\vec{E};\quad
\vec{\mu}=-{\delta H_{eff}\over \delta\vec{H}}=4\pi\beta_M\vec{H}
\end{equation}
where $\vec{p},\,\vec{\mu}$ are the electric, magnetic dipole moments
generated under the influence of external electric, magnetizing fields $\vec{E},
\,\vec{H}$.  We recognize $\alpha_E,\,\beta_M$ as being the the electric,
magnetic polarizabilities, which have obvious and intutive classical 
meanings\cite{ho}
\begin{itemize}
\item [i)] in the presence of an external electric field the positive, negative
components of the charge distribution move in opposite directions, resulting
in an induced electric dipole moment;
\item [ii)] in the presence of an external magnetizing field intrinsic 
magnetic moments tend to align, creating a paramagnetic effect, while 
orbital moments generate a diamagnetic component, in accord with Lenz law.
\end{itemize}
Using the above effective Hamiltonians the
Compton scattering amplitude becomes
\begin{equation}
{\rm Amp}^{(2)}=\hat{\epsilon}\cdot\hat{\epsilon}'\left({-e^2\over
M}+\omega\omega' \; 4\pi\alpha_E^p \right)
+\hat{\epsilon}\times\vec{k}\cdot\hat{\epsilon}'
\times\vec{k}'\;4\pi\beta_M^p+\; {\cal O}(\omega^4) \; .
\end{equation}
and the resultant differential scattering cross section is 
\begin{eqnarray}
{d\sigma\over d\Omega}&=&\left({\alpha\over M}\right)^2
\left({\omega'\over \omega}\right)^2\left[{1\over 2}
(1+\cos^2\theta)\right.\nonumber\\
&-&\left.{ M\omega\omega'\over \alpha}\left({1\over 2}
(\alpha_E^p+\beta_M^p)(1+\cos\theta)^2+{1\over
2}(\alpha_E^p-\beta_M^p)(1-\cos\theta)^2\right)+\ldots\right],
\nonumber\\
\quad
\end{eqnarray}
where $\alpha=e^2/4\pi$ is the fine structure constant.  It is clear then that 
$\alpha_E,\,\beta_M$ can be extracted via careful measurement of the 
differential cross section and previous experiements at SAL and MAMI have
yielded the values\cite{nat}\footnote{In order to put these 
numbers in perspective
note that for a hydrogen atom one finds $\alpha_E(H)\sim$ Volume($H$) 
while for 
the proton Eq. \ref{eq:yy} gives $\alpha_E(p)\sim 10^{-3}$Volume($p$), so that
the proton is a much more strongly bound system.} 
\begin{equation}
\alpha_E^p=(12.1\pm 0.8\pm 0.5)\times
10^{-4}\,\,{\rm fm}^3;\quad\beta_M^p=(2.1\mp 0.8\mp 0.5)\times
10^{-4}\,\,{\rm fm}^3.\label{eq:yy}
\end{equation}
At this meeting Wissmann announced new values obtained from precise
$p(\gamma,\gamma)p$ measurements using the TAPS and LARA spectrometers\cite{wis}
\begin{equation}
\alpha_E^p=(12.24\pm 0.24\pm 0.54)\times
10^{-4}\,\,{\rm fm}^3;\quad\beta_M^p=(1.57\mp 0.24\mp 0.54)\times
10^{-4}\,\,{\rm fm}^3.
\end{equation}
These measured numbers can be compared to the corresponding quantities
calculated in heavy baryon chiral perturbation theory (HB$\chi$pt)
 at ${\cal O}(p^3)$\cite{BKKM92}
\begin{equation}
\alpha_E^p=10K_p=12.7\times 10^{-4}\,\,{\rm fm}^3,\quad
\beta_M^p=K_p=1.3\times 10^{-4}\,\,{\rm fm}^3
\end{equation}
where $K_p=\alpha g_A^2/192\pi F_\pi^2m_\pi$.  Here $g_A\simeq 1.266$
is the axial coupling constant in neutron beta decay and $F_\pi\simeq
92.4$ MeV  is the pion decay constant. 
Of course, one must include higher order terms in order to properly judge the
convergence behavior of the series, and such a calculation at ${\cal O}(p^4)$ 
has been performed by Bernard, Kaiser,
Schmidt and Mei\ss ner (BKSM)\cite{BKSM93}.  At this order counterterms are
required, which were
estimated by BKSM by treating higher 
resonances---including $\Delta$(1232)---as very heavy
with respect to the nucleon, yielding
\begin{equation}
\alpha_E^p=(10.5\pm2.0)\times
10^{-4}\,\,{\rm fm}^3;\quad\beta_M^p=(3.5\pm 3.6)\times
10^{-4}\,\,{\rm fm}^3
\end{equation}
where the uncertainty is associated with the counterterm contribution
from the $\Delta(1232)$ and from $K,\eta$ loop effects.  Agreement remains 
good between theory and experiment, supporting the view that the
pion cloud gives a good description of such quantities.

The above results are well known and our task today is to extend 
this discussion
to indlude spin degrees of freedom.  In this case the general Compton 
amplitude can written in the general form
\begin{eqnarray}
T &=& A_1(\omega,z)\vec{\epsilon}^{\, \prime}\cdot\vec{\epsilon}
+A_2(\omega,z)\vec{\epsilon}^{\,  \prime}\cdot\hat{k} \; \vec{\epsilon}
\cdot\hat{k}^\prime \nonumber\\
&+&iA_3(\omega,z)\vec{\sigma}\cdot(\vec{\epsilon}^{\, \prime}\times
\vec{\epsilon})
+iA_4(\omega,z)\vec{\sigma}\cdot(\hat{k}^\prime \times\hat{k})
\vec{\epsilon}^{\,  \prime} \cdot\vec{\epsilon} \nonumber\\
&+& iA_5(\omega,z)\vec{\sigma}\cdot[(\vec{\epsilon}^{\,  \prime} \times
\hat{k}) \vec{\epsilon}\cdot\hat{k}^\prime -(\vec{\epsilon}\times
\hat{k}^\prime ) \vec{\epsilon}^{\,  \prime} \cdot\hat{k}]\nonumber\\
&+& iA_6(\omega,z)\vec{\sigma}\cdot[(\vec{\epsilon}^{\, \prime}\times
\hat{k}^\prime ) \hat{\epsilon}\cdot\hat{k}^\prime -(\vec{\epsilon}\times
\hat{k})\vec{\epsilon}^{\, \prime} \cdot\hat{k}],
\end{eqnarray}
and each amplitude can be expanded in terms of a lowest order Born 
contribution plus a higher order and structure-dependent polarizability
term.  In the case of the spin-dependent amplitudes $A_{3,4,5,6}$ such
structure effects arise at ${\cal O}(\omega^3)$ and can be characterized in
terms of an effective Hamiltonian involving four "spin-polarizabilities"
\begin{equation}
H_{eff}^{(3)}=-{1\over 2}4\pi(\gamma_{E1}^p\vec{\sigma}\cdot\vec{E}\times
\dot{\vec{E}}+\gamma_{M1}^p\vec{\sigma}\cdot\vec{H}\times
\dot{\vec{H}}-2\gamma_{E2}^pE_{ij}\sigma_iH_j+2\gamma_{M2}^pH_{ij}\sigma_iE_j)
\label{eq:mno}
\end{equation} 
where 
\begin{equation}
E_{ij}={1\over 2}(\nabla_iE_j+\nabla_jE_i),\quad H_{ij}={1\over 2}(\nabla_iH_j
+\nabla_jH_i)
\end{equation}
denote electric and magnetizing field gradients.  While these quantities
are mathematically well-defined via Eq. \ref{eq:mno}, I am unable to 
provide a good
physical picture.  The parameters $\gamma_{E1},\,
\gamma_{M1}$ are similar to the classical Faraday rotation, wherein the
linear polarization of the photon passing longitudinally through a 
magnetized medium exhibits a rotation due to the difference in index of
refraction for photons with circular polarization parallel and antiparallel
to the direction of magnetization.  However, I don't know how to go much farther
than this and will offer a bottle of fine Mainz Kupfenberg Sekt to anyone
who is able to provide me such a classical picture.

On the theoretical side the chiral predictions for the spin-polarizabilities 
at ${\cal O}(p^3)$ are given by\cite{bkm},\cite{hhk}
\begin{equation}
\gamma_{E1}^p=-{10K_p\over \pi m_\pi},\quad\gamma_{M1}^p=-{2K_p\over \pi 
m_\pi},\quad\gamma_{E2}^p=
{2K_p\over \pi m_\pi},\quad\gamma_{M2}^p={2K_p\over \pi m_\pi}
\end{equation}
and extensions to ${\cal O}(p^4)$, necessary to assess the convergence, were
presented to this workshop by Hemmert\cite{hem} and by McGovern\cite{mcg}.  
While the numerical 
results of the two calculations are in agreement, the interpretation in 
terms of polarizabilities is presently in dispute.  The problem arises from
HB$\chi$pt diagrams wherein there is a nucleon pole on one side of which
exists a pion loop renormalizing an electromagnetic vertex.  In ordinary
relativistic perturbation theory such diagrams are clearly one particle 
reducible and would be discarded as having nothing
to do with polarizabilities.  However, this is not so clear in the heavy baryon
chiral calculation and this is where the problem lies at present.  Ulf
Meissner has promised a resolution in $n$-weeks but has not yet given a 
definitive value (or even a bound!) for the number $n$.  In the meantime I
shall quote the J\"{u}lich calculations
\begin{eqnarray}
\gamma_{E1}^p&=&-{10K_p\over \pi m_\pi}(1-{29\over 20}{\pi m_\pi\over M})
\nonumber\\
\gamma_{M1}^p&=&-{2K_p\over \pi m_\pi}(1-{11\over 4}{\pi m_\pi\over M})
\nonumber\\
\gamma_{E2}^p&=&{2K_p\over \pi m_\pi}(1-{2\kappa_n+3\over 4}
{\pi m_\pi\over M})\nonumber\\
\gamma_{M2}^p&=&{2K_p\over \pi m_\pi}(1-{3\over 4}{\pi m_\pi\over M})
\label{eq:sp}
\end{eqnarray}

On the experimental side, there exist as yet no direct polarized Compton
scattering measurements.  However, a global analysis of unpolarized Compton
data by the LEGS group has yielded the value\cite{ton}\footnote{Note here 
that we have subtracted the pion pole contribution.}
\begin{equation}
\gamma_\pi=-\gamma_{E1}-\gamma_{M2}+\gamma_{E2}+\gamma_{M1}=
(15.7\pm 2.3\pm 2.8\pm 2.4)\times 10^{-4}\,{\rm fm}^4
\end{equation} 
in disagreement with the theoretical prediction 
\begin{equation}
\gamma_\pi={2K_p\over \pi m_\pi}(4-(\kappa_n+{9\over 2}){\pi
m_\pi\over M})=3.3\times 10^{-4}\,{\rm fm}^{4}
\end{equation}
from Eq. \ref{eq:sp}.  However, Wissman has announced a new value from the TAPS
data
\begin{equation}
\gamma_\pi=(7.4\pm 2.3)\times 10^{-4}\,{\rm fm}^4
\end{equation}
which is in better agreement with theory.  The 
other quantity about which much has been written is the forward
spin polarizability $\gamma_0$, which is given by the first moment
of the DGH sum rule
\begin{equation}
\gamma_0=\gamma_{E1}+\gamma_{M2}+\gamma_{E2}+\gamma_{M1}=\int_{\omega_0}^\infty
{d\omega\over \omega^3}(\sigma_{3\over 2}(\omega)-\sigma_{1\over 2}(\omega))
\end{equation}
Drechsel at this meeting has quoted perhaps the best current value of the
sum rule, based upon the MAID analysis,
\begin{equation}
\gamma_0=-0.80\times 10^{-4}\,{\rm fm}^4
\end{equation}
which is in reasonable agreement with previous determinations.

While at present we do not have direct experimental values for the four
spin-polarizabilities, Barbara Pasquini described a way by which they can be
obtained using a dispersive analysis of the Compton process\cite{pas}.  
One assumes
that the Compton amplitudes $A_i$ can be
represented in terms of once subtracted dispersion relations at fixed t
\begin{equation}
A_i(\nu,t)=A_i^{\rm Born}(\nu,t)+(A_i(0,t)-A_i^{\rm Born}(0,t))
+{2\nu^2\over \pi}P\int_{\nu_{thr}}^\infty{{\rm Im}A_i(\nu',t)\over
\nu' ({\nu'}^2-\nu^2)}.
\end{equation}
Here Im$A_i(\nu',t)$ is evaluated using empirical photoproduction data
while the subtraction constant $A_i(0,t)-A_i^{\rm Born}(0,t)$ is
represented via use of t-channel dispersion relations
\begin{equation}
A_i(0,t)-A_i^{\rm Born}(0,t)=a_i+a_i^{t-pole}+
{t\over \pi}\left(\int_{4 m_\pi^2}^\infty
-\int_{-\infty}^{- 4 M m_\pi - 2 m_\pi^2} dt'
{{\rm Im}_tA_i(0,t')\over t'(t'-t)}\right)
\end{equation}
with Im$_tA_i$ evaluated using the contribution from the
$\pi\pi$ intermediate state.  In principle then there remain six
unknown subtraction constants $a_i$ to be determined
empirically.  However, in view of the limitations posed by the the data,
Drechsel et al. note that four of these quantities can be reasonably
assumed to obey unsubtracted forward dispersion relations, while the remaining
two---$\alpha_E-\beta_M$ and $\gamma_\pi$---can be treated as
parameters and fitted from the data.  Once this is done the other
spin polarizabilities may be
extracted using sum rules, as done above in the case of the forward
spin polarizability.  The results of this process
are compared in Table 1 with predictions of chiral perturbation theory and
one finds generally satisfactory agreement.  

\begin{table}
\begin{center}
\begin{tabular}{ccc}
polarizability& HB$\chi$pt & Dispersive Evaluation\\
$\gamma_{E1}^p$&-1.8&-4.4\\
$\gamma_{M1}^p$&2.9&2.9\\
$\gamma_{E2}^p$&1.8&2.2\\
$\gamma_{M2}^p$&0.7&0.0
\end{tabular}
\caption{Calculated and "experimental" values for spin polarizabilities 
obtained via dispersion relations.  All are in units of $10^{-4}$ fm$^4$.}
\end{center}
\end{table}

It has been noted by Babusci et al.\cite{bab} and by Holstein et al.\cite{hol}
that one can
extend this analysis to include terms of ${\cal O}(\omega^4)$ in the Compton
amplitude by introducing higher order polarizabilities via  
\begin{equation}
H_{eff}^{(4)}=-{1\over 2}4\pi\alpha_{E\nu}^p\dot{\vec{E}}^2
-{1\over 2}4\pi\beta_{M\nu}^p\dot{\vec{H}}^2
-{1\over 12}4\pi\alpha_{E2}^pE_{ij}^2
-{1\over 12}4\pi\beta_{M2}^p H_{ij}^2\label{eq:mmm}
\end{equation} 
Likewise Holstein et al. have extended this to ${\cal O}(\omega^5)$ by
defining higher order spin-polarizabilities---
\begin{eqnarray}
H_{eff}^{(5)}&=&-{1\over 2}4\pi\left[\gamma_{E1\nu}^p\vec{\sigma}\cdot\dot{\vec{E}}
\times\ddot{\vec{E}}+\gamma_{M1\nu}^p\vec{\sigma}\cdot\dot{\vec{H}}\times
\ddot{\vec{H}}-2\gamma_{E2\nu}^p\sigma_i\dot{E}_{ij}\dot{H}_j+2\gamma_{M2\nu}^p
\sigma_i\dot{H}_{ij}\dot{E_j}\right.\nonumber\\
&+&\left.4\gamma_{ET}^p\epsilon_{ijk}\sigma_iE_{j\ell}\dot{E}_{k\ell}
+4\gamma_{MT}^p\epsilon_{ijk}\sigma_iH_{j\ell}\dot{H}_{k\ell}
-6\gamma_{E3}^p\sigma_iE_{ijk}H_{jk}+6\gamma_{M3}^p\sigma_iH_{ijk}E_{jk}\right]\nonumber\\
\quad 
\end{eqnarray}
where 
\begin{eqnarray}
{(E,H)}_{ijk}&=&{1\over 3}(\nabla_i\nabla_j{(E,H)}_k+\nabla_i\nabla_k{(E,H)}_j
+\nabla_j\nabla_k{(E,H)}_i)\nonumber\\
&-&{1\over
15}(\delta_{ij}\nabla^2(E,H)_k+\delta_{jk}
\nabla^2(E,H)_i+\delta_{ik}\nabla^2(E,H)_j)
\end{eqnarray}
are the (spherical) tensor gradients of the electric and magnetizing 
fields.  Each of these new higher order polarizabilities 
can be extracted via sum
rules from the Mainz dispersive analysis and results are compared with 
chiral predictions in Table 2.  Agreement is obviously quite
satisfactory, except for $\alpha_{E\nu}$. 
Despite the success of this program it would be highly desirable to 
measure such quantities directly and Wissmann has suggested a 
$\vec{p}(\vec{\gamma},
\gamma)p$ program by which it might be possible to achieve this.

\begin{table}
\begin{center}
\begin{tabular}{ccc}
polarizability&HB$\chi$pt&Dispersive value\\
$\alpha_{E\nu}^p$&2.4&-3.8\\
$\beta_{M\nu}^p$&7.5&9.3\\
$\alpha_{E2}^p$&22.1&29.3\\
$\beta_{M2}^p$&-9.5&-24.3\\
$\gamma_{E1\nu}^p$&-2.4&-3.4\\
$\gamma_{M1\nu}^p$&1.8&2.2\\
$\gamma_{E2\nu}^p$&1.6&1.3\\
$\gamma_{M2\nu}^p$&-0.1&-0.6
\end{tabular}
\caption{Calculated and "experimental" values for higher order 
polarizabilities obtained via dispersion relations.  Spin independent and
spin dependent polarizabilities are in units of $10^{-4}$ fm$^5$ and
$10^{-4}$ fm$^6$ respectively}
\end{center}
\end{table}

\section{Virtual Compton Scattering}

Nicole d'Hose discussed the virtual Compton scattering (VCS) process by
which one can measure "generalized" (q-dependent) polarizabilities\cite{dho}.
In order
to understand the meaning of such quantities, recall that in ordinary electron
scattering measurement of the q-dependent charge form factor allows access,
via Fourier transform, to the nucleon charge {\it density}.  In an analogous
fashion measurement of a generalized polarizability such as $\alpha_E(q)$
permits one to determine the polarization density of the nucleon.  On the 
experimental side this is an extremently challenging process because the 
generalized polarizabilities can be determined only after (large) Bethe-Heitler
and Born diagram contributions have been subtracted.\footnote{Radiative
corrections are also substantial here.}  One then seeks a systematic 
deviation growing with $\omega'$ of the measured cross section from that 
predicted with only Bethe-Heitler plus Born input in order to extract the
desired signal.  This has been achieved in a recent MAMI experiment\cite{roche} and what 
results is information on the two combinations
\begin{eqnarray}
P_{LL}-{1\over \epsilon}P_{TT}&=&a_0\alpha_E(q)-c_1\gamma_{M2}(q)
+c_2M^{M1-M1}(q)\nonumber\\
P_{LT}&=&b_0\beta_M(q)+c_3M^{C0-M1}(q)-c_4\gamma_{E2}(q)
\end{eqnarray}
where here the multipoles $M^{M1-M1},\,M^{C0-M1}$ have no RCS
analogs.  The extracted numbers are given in Table 3 together with values
calculated in various models as well as in HB$\chi$pt.  It is remarkable
that once again agreement with the simple chiral calculation is outstanding,
despite the fact that the measurement took place at q=0.6 GeV, where one
should question the validity of the chiral approach.

\begin{table}
\begin{center}
\begin{tabular}{c|c|c}
 &$P_{LL}-P_{TT}/\epsilon$&$P_{LT}$\\
\hline\\
expt.& $23.7\pm 2.2\pm 0.6\pm 4.3$& $-5.0\pm 0.8\pm 1.1\pm 1.4$\\
HB$\chi$pt\cite{hhks}&26.0&-5.3\\
L$\sigma$M\cite{lsm}&11.5&0.0\\
ELM\cite{elm}&5.9&-1.9\\
NRQM\cite{nrq}&11.1&-3.5
\end{tabular}
\caption{Measured and calculated values for generalized polarizabilities.
All are in units of GeV$^{-2}$.}
\end{center}
\end{table}

On the theoretical side, Marc Vanderhaeghen reported on an extension of the
Mainz dispersive RCS analysis to the case of VCS\cite{vdh}.  
This extension is not
as straightforward as it might appear, since replacement of a real photon
by a virtual one requires now {\it twelve} invariant amplitudes which are
functions of three variables, which may be taken as $\omega,\theta,Q^2$.
Analysis of the asymptotic dependence is correspondingly much more complex.  
The calculation is still in progress but Vanderhaeghen presented 
preliminary results
for four of the generalized polarizabilities compard to the chiral 
predictions.   Agreement is satisfactory though not outstanding, but Hemmert
argued that ${\cal O}(p^4)$ corrections will improve matters.  

\section{Conclusion}

The subject of polarizabilities---both spin-dependent as well as 
spin-independent---in both RCS and VCS is undergoing a major emphasis
due to the presence of high quality data from machines such as MAMI,
Bates, SAL,
CEBAF, etc. and we have heard lots of exciting results at this meeting.
I believe that at GDH2002 we can anticipate equally exciting 
developments.  For RCS:
\begin{itemize}
\item [i)] theory---definitive ${\cal O}(p^4)$ calculation of all
polarizabilities in HB$\chi$pt will be completed;
\item [ii)] experiment---analysis of the MAMI results will be completed 
and preliminary $\vec{p}(\vec{\gamma},\gamma)p$ data will be presented.
\end{itemize}
For VCS:
\begin{itemize}
\item [i)] theory---existing HB$\chi$pt calculations will be extended to 
${\cal O}(p^4)$;
\item [ii)] experiment---analysis of the MAMI experiment will be complete 
and results from JLab and Bates experiments will be presented.  Preliminary
$\vec{p}(\vec{\gamma}^*,\gamma)p$ data will be available. 
\end{itemize}
These are exciting times for low energy Compton scattering and I look forward
to our meeting two years hence in Italy.

\begin{center}
{\bf Acknowledgement}
\end{center}
It is a pleasure to acknowledge the hospitality of the organizers of this
workshop.  This work was supported in part by the National Science Foundation.
\\

\end{document}